\documentclass[prl,preprint, superscriptaddress]{revtex4-1}

\usepackage[dvipdfmx]{graphicx}
\usepackage{dcolumn}
\usepackage{bm}

\usepackage{color}

\begin{document}


\title{Giant Rashba system on a semiconductor substrate with tunable Fermi level: Bi/GaSb(110)-(2$\times$1)}

\author{Takuto Nakamura}
\affiliation{Department of Physics, Graduate School of Science, Osaka University, Toyonaka 560-0043, Japan}
\author{Yoshiyuki Ohtsubo}
\email{y\_oh@fbs.osaka-u.ac.jp}
\affiliation{Department of Physics, Graduate School of Science, Osaka University, Toyonaka 560-0043, Japan}
\affiliation{Graduate School of Frontier Biosciences, Osaka University, Suita 565-0871, Japan}
\author{Naoki Tokumasu}
\affiliation{Department of Physics, Graduate School of Science, Osaka University, Toyonaka 560-0043, Japan}
\author{Patrick Le F\`evre}
\author{Fran\c{c}ois Bertran}
\affiliation{Synchrotron SOLEIL, Saint-Aubin-BP 48, F-91192 Gif sur Yvette, France}
\author{Shin-ichiro Ideta}
\author{Kiyohisa Tanaka}
\affiliation{Institute for Molecular Science, Okazaki 444-8585, Japan}
\author{Kenta Kuroda}
\author{Koichiro Yaji}
\author{Ayumi Harasawa}
\author{Shik Shin}
\author{Fumio Komori}
\affiliation{Institute for Solid State Physics, The University of Tokyo, 5-1-5 Kashiwanoha, Kashiwa, Chiba 277-8581, Japan}
\author{Shin-ichi Kimura}
\email{kimura@fbs.osaka-u.ac.jp}
\affiliation{Department of Physics, Graduate School of Science, Osaka University, Toyonaka 560-0043, Japan}
\affiliation{Graduate School of Frontier Biosciences, Osaka University, Suita 565-0871, Japan}

\date{\today}

\begin{abstract}
We fabricated spin-polarized surface electronic states with tunable Fermi level from semiconductor to low-dimensional metal in the Bi/GaSb(110)-(2$\times$1) surface using angle-resolved photoelectron spectroscopy (ARPES) and spin-resolved ARPES.
The spin-polarized surface band of Bi/GaSb(110) exhibits quasi-one-dimensional character with the Rashba parameter $\alpha _{\rm R}$ of 4.1 and 2.6 eV\AA \ at the $\bar{\Gamma}$ and $\bar{\rm Y}$ points of the surface Brillouin zone, respectively.
The Fermi level of the surface electronic state is tuned in situ by element-selective Ar-ion sputtering on the GaSb substrate.
The giant Rashba-type spin splitting with switchable metallic/semiconducting character on semiconductor substrate makes this system a promising candidate for future researches in low-dimensional spintronic phenomena.
\end{abstract}

\maketitle

\section{Introduction}
Spin-polarized conduction path in low-dimensional systems (spintronic circuits), which can be realized by the surface states of topological insulators (TIs) \cite {Hasan10} for example, has garnered significant attention recently, because it hosts exotic spintronic phenomena, such as spin-charge conversion and unconventional terahertz emission \cite{Manchon15, Han18, Matthias18}.
Among them, the Rashba effect at crystal surfaces is a promising platform for designing such low-dimensional systems \cite {Rashba}.
Due to the lack of the space inversion symmetry at the surface, spin-orbit interaction (SOI) induces a sizable spin polarization in surface electronic states even without ferromagnetic materials.
The conduction path from surface electronic states with Rashba-type SOI is intrinsically switchable from semiconductor to metal, in contrast to the topologically protected, always metallic surface states of TIs, which is a great strongpoint for designated spintronic circuits.
Moreover, as such Rashba-type SOI can occur in surface systems fabricated on ordinary semiconductor substrates \cite{Gierz09, Yaji10}, it has been studied extensively for the development of sptronic devices, which exhibit excellent compatibility with semiconductor substrates.

Recently, we reported that a quasi-one-dimensional (Q1D) surface, Bi chains on a semiconductor InAs(110) substrate, exhibits giant Rashba-type spin splitting in its electronic states \cite {Nakamura18}.
Here, we obtained the Rashba parameter $\alpha _{\rm R}$, a widely used scale of Rashba-type SOI, as 5.5 eV {\AA}, which is close to the maximum value reported for Rashba-like systems \cite{Nakagawa07, Ast07, Ishizaka11, Miyamoto14, Liebmann16}.
The Bi Q1D chain did not become a metal but a semiconductor with an energy gap of 0.04 eV between the Fermi level ($E_{\rm F}$) and the valence band maximum at the $\bar{\rm Y}$ point, which limits its application for spintronic circuits. This issue can be resolved by tuning $E_{\rm F}$ to make the surface state metallic so that spin-polarized transport can be realized.
To achieve this $E_{\rm F}$ tuning of surface electronic structure, some methods have been reported, such as co-evaporation with the neighboring element in the periodic table \cite {Ast08}, alkali-metal evaporation \cite{AM}, and excess evaporation of the elements to form a surface superstructure for providing electrons to the surface electronic states \cite{Ohtsubo12, Sakamoto13, Nakatsuji13}.
Application of these methods to the Bi Q1D chains can be fruitful for their implementation in spintronic devices.
In parallel, it is also desirable to increase the energy gap of the substrate compared with that of InAs (0.36 eV at 300 K), which is too small to ignore the thermal spin-degenerate carriers from substrates, while maintaining a similar surface electronic structure.

In this work, we investigate the surface electronic states of the zig-zag chains of Bi on GaSb(110), which exhibit giant Rashba-type spin splitting. The surface atomic structure is depicted in Fig. 1(a), which is similar to that of Bi/InAs(110) reported previously; the bulk bandgap of GaSb (0.68 eV at 300 K) is about two times larger than that of InAs.
The Q1D feature of the surface bands of Bi/GaSb(110) was demonstrated by angle-resolved photoelectron spectroscopy (ARPES).
We obtained $\alpha _{\rm R}$ as 4.1 and 2.6 eV {\AA} at the $\bar{\Gamma}$ and $\bar{\rm Y}$ points of the surface Brillouin zone (SBZ), respectively. These values are slightly smaller than those of Bi/InAs(110), but are still quite a large value.
Further, we tuned $E_{\rm F}$ of Bi/GaSb(110) by element-selective Ar-ion sputtering on the GaSb substrates.
The surface bands rigidly shifted by $\sim$100 meV toward the lower binding energy; consequently, these bands crossed $E_{\rm F}$, indicating the in-situ transition of surface electronic states from semiconductor to low-dimensional and spin-polarized metal.

\section{Experimental Methods}

The clean GaSb(110)-(1$\times$1) substrates were fabricated by two different methods. In the first method, a side face of single-crystal GaSb(001) wafers was cleaved. The second method involved repeated cycles of Ar-ion sputtering (with acceleration energy of 0.5 - 1.0 keV) and annealing at temperatures up to 700 K. The prepared samples are denoted as ``cleaved'' and ``S$\&$A'' surfaces, respectively.
A few monolayers of Bi were evaporated from a Knudsen cell at room temperature and then annealed at 550 K for more than 10 minutes. The atomic structures of the obtained surfaces were visualized by low-energy electron diffraction (LEED). The LEED patterns for both the samples are shown in Figs. 1(b) and 1(c), which exhibit sharp patterns with low background (2$\times$1) periodicity, indicating well-defined surface of Bi/GaSb(110). The distortions in these patterns are due to the flat micro channel plate used for LEED measurements.
The common (2$\times$1) patterns in both the cases suggest the formation of Bi Q1D atomic chains on GaSb(110) substrates \cite {Gemmeren98}.

ARPES and core-level photoelectron spectroscopy measurements were performed with a He discharge lamp and synchrotron radiation at the CASSIOP\'EE beamline of synchrotron SOLEIL.
Spin-resolved ARPES (SARPES) measurements were conducted at the Institute for Solid State Physics, the University of Tokyo with linearly polarized laser source ($h\nu$ = 6.994 eV) \cite {Yaji16}.
The energy resolution and the position of $E_{\rm F}$ were calibrated by the Fermi edge of Ta foils attached to the samples.
For the ARPES (SARPES) measurements, the incident plane of photons was [001] ([$\bar{1}$10]), and the energy resolution was 12 meV (30 meV).
The effective Sherman function of the SARPES spin detector was set to 0.27.


\section{Results and Discussion}
Figures 2(a) and 2(c) show the ARPES intensity plots of the Bi-induced (2$\times$1) surface on the cleaved GaSb(110) substrates (called cleaved Bi/GaSb) along $\bar{\Gamma}$--$\bar{\rm X}$ and $\bar{\rm Y}$--$\bar{\rm M}$, respectively, in the SBZ shown in Fig. 2(e).
Paired parabolic bands are located slightly below $E_{\rm F}$ at around $\bar{\Gamma}$ and $\bar{\rm Y}$, demonstrating semiconducting characteristics of the surface.
The shapes of these bands are qualitatively similar to those of the Q1D surface electronic states of Bi/InAs(110)-(2$\times$1) \cite {Nakamura18}, which originate from the common $``missing-row''$ model with the zig-zag Bi chains, as shown in Fig. 1(a) \cite{Gemmeren98, Betti99}.

Figures 2(b) and 2(d) show the ARPES band dispersions of Bi/GaSb(110)-(2$\times$1) fabricated on the S$\&$A GaSb(110) substrates (called S$\&$A Bi/GaSb hereafter).
The obtained band dispersions are similar to those observed for the cleaved Bi/GaSb, but they are shifted toward lower binding energies.
Especially, the top of the valence band along $\bar{\rm Y}$--$\bar{\rm M}$ disappears, which suggests that the band crosses $E_{\rm F}$.
The right side of Fig. 2(f) shows the constant-energy contour of the S$\&$A Bi/GaSb surface with the binding energy range of 0 $\pm$ 20 meV, which corresponds to a Fermi surface.
A butterfly-like, highly anisotropic Fermi surface appears near $\bar{\rm Y}$, which is consistent with the constant-energy contour of the cleaved Bi/GaSb surface at the binding energy of 95 $\pm$ 20 meV.
The detailed discussion on this point is shown in the following part.
Comparing the ARPES intensity plots from the cleaved and S$\&$A surfaces in Figs. 2(a-d), the line width and background of the S$\&$A sample are slightly broader and higher than those of the cleaved ones.
It would be due to the damages by the Ar-ion sputtering.
The same trend can be also found in the LEED patterns (Figs. 1 (b) and 1(c))

The pair of parabolic bands are symmetric with respect to the time-reversal-invariant momenta ($\bar{\Gamma}$ and $\bar{\rm Y}$), which is consistent with that expected by the conventional Rashba effect \cite {Rashba}.
Figures 3(a) and 3(b) show the SARPES (upper figure) and ARPES (lower figure) momentum distribution curves (MDCs) along $\bar{\rm Y}$--$\bar{\rm M}$ at the binding energies of 20 meV and 100 meV, respectively.
The filled (open) triangles in the upper figures indicate the spin-resolved MDCs parallel (anti-parallel) to [001].
The peak positions in the upper and lower figures remain same despite the different photon energy, suggesting that these peaks originate from the surface states.
In the MDCs, peaks appear at $\pm$0.02 and $\pm$0.10 \AA $^{-1}$ corresponding to the binding energy of 20 meV (at $\pm$0.14 \AA $^{-1}$ and 100 meV), which agree with the band dispersions shown in Fig. 2(d). These MDC peaks are clearly spin polarized.
The spin-polarization orientation of these peaks is reversed together with the sign of $k_{//[\bar{1}10]}$, as expected in the Rashba-type spin-orbit splitting of surface bands.
These spin polarizations are consistent with the spin-polarized surface electronic structure of Bi/InAs(110)-(2$\times$1) \cite {Nakamura18}.

The magnitude of Rashba-type spin splitting is often characterized by the Rashba parameter $\alpha_{\rm R}$ = 2$E_{\rm R}$/$k_0$, assuming parabolic dispersion.
Here, $E_{\rm R}$ and $k_0$ are the difference in energy and wavevector between the Kramers degenerate point and the top (or bottom) of the parabolic band, respectively.
From the surface band dispersions in Figs. 2(a) and 2(b), $E_{\rm R}$ and $k_0$ along $\bar{\Gamma}$--$\bar{\rm X}$ ($\bar{\rm Y}$--$\bar{\rm M}$) can be obtained as 0.29 (0.065) eV and 0.14 (0.05) \AA, respectively.
Accordingly, $\alpha_{\rm R}$ along $\bar{\Gamma}$--$\bar{\rm X}$ ($\bar{\rm Y}$--$\bar{\rm M}$) is 4.1 (2.6) eV\AA.
Although the values of $\alpha _ { \rm R}$ are slightly smaller than that obtained for Bi/InAs(110)-(2$\times$1) with a similar surface atomic structure (5.5 eV {\AA} along $\bar{\Gamma}$--$\bar{\rm X}$ and 3.6 eV {\AA} along $\bar{\rm Y}$--$\bar{\rm M}$) \cite{Nakamura18}, they are still three-four times larger than that reported for other Q1D and 1D systems such as Pt/Si(110)\cite {Park13} and the edges of Bi(111)/Si(111) \cite {Takayama15}, and are comparable to the values obtained for giant Rashba systems such as two-dimensional Bi/Ag surface and three-dimensional BiTeI \cite{Nakagawa07, Ast07, Ishizaka11}.

The difference of $\alpha _ { \rm R}$ between Bi/GaSb(110) and Bi/InAs(110) can be attributed to the slight variation in the surface atomic structure.
Table I shows the bond lengths and relative atomic displacements along [110] plane of the Bi atomic chains on InAs(110) and GaSb(110) \cite {Lottermoser98, Gemmeren98, Gay00, Betti99}.
The bulk lattice constants of InAs and GaSb are 6.06 \AA\ and 6.10 \AA, respectively, with a difference of less than 1 $\%$.
Compared to this small difference, the inter-atomic distances in the Bi-adsorbed surfaces are large.
Bond lengths between Bi atoms of Bi/GaSb are 3-4 \% larger. In contrast, the bond lengths of Bi-Ga and Bi-Sb in Bi/GaSb are shorter than those of Bi-In and Bi-As in Bi/InAs (difference of 4-6 \%), except for the bond Bi1-Ga1 (In1).
In addition, the vertical displacements of the Bi atomic chains in Bi/GaSb(110) are larger than that in Bi/InAs(110) by 0.05 \AA\ for $\Delta_{1-2}$ and 0.15 \AA\ for $\Delta_{3-4}$.
This difference may be attributed to the ionicity of III$-$V semiconductor, as reported in an earlier work \cite {Betti99}.
Actually, the Phillips ionicity of GaSb (0.26) is smaller than that of InAs (0.36) \cite {Phillips70}.
Thus, the ionic character of the surface anions and cations can affect the valence bond of Bi atoms and modify the surface bond lengths.
It is well established the magnitude of Rashba-type spin splitting is strongly influenced by the orbital composition of the surface states and the charge distribution near the surface atoms, and both of them can be modified by the structural and valence changes around surface Bi atoms as discussed above \cite{Nagano09, Ishida14}.
Therefore, this can be rational cause for the change in the magnitude of Rashba effect in Bi/GaSb(110) and Bi/InAs(110).

Further insights into the dependence of the surface cleaning method on the surface-band shift can be obtained from the left-side figures in Figs. 3(c) and 3(d), which show the peak positions of the energy distribution curves (EDCs) from the ARPES intensity plots in Figs. 2(b,d) ($\bar{\Gamma}$--$\bar{\rm X}$) and 2(c,e) ($\bar{\rm Y}$--$\bar{\rm M}$), respectively.
The circles (squares) indicate the EDC peaks of cleaved (S$\&$A) Bi/GaSb.
In the right-side figures, the peak positions of S$\&$A Bi/GaSb are shifted by 95 meV and become consistent with those of cleaved Bi/GaSb. This suggests that the S$\&$A operation affects the rigid band shift without causing any deformation to the band dispersion.
This can be confirmed by the constant-energy contours shown in Fig. 2(f), which indicate that the S$\&$A Bi/GaSb at 0 $\pm$ 20 meV is consistent with that of cleaved Bi/GaSb at 95 $\pm$ 20 meV.
The only difference is that the center of the batterfly-shape contour of the cleaved Bi/GaSb has high intensity.
It would be from the high-binding-energy tail of the surface bands lying close to $E_{\rm F}$ (around the apex of the paired parabola) which is located in the unoccupied region in S$\&$A Bi/GaSb.

We examined the variation of core-level spectra to gain further information on the origin of this rigid band shift in S$\&$A Bi/GaSb.
The core-level spectra of cleaved and S$\&$A samples obtained with 40.8-eV and 80-eV photons are shown in Figs. 4 and 5, respectively.
Figures 4(a-c) and 4(d-g) show the spectra of GaSb(110) and Bi/GaSb(110)-(2$\times$1), respectively.
Figures 4(a) and 4(d) illustrate the wide-range spectra measured by He-II light source ($h\nu$ = 40.8 eV). The spectra in Figs. 4(b,c) and 4(d-f) are obtained after subtracting the Shirley-type background from Figs. 4(a) and 4(d), respectively.
Further annealing for more than 10 minutes in the S$\&$A samples caused no change on the area of Bi 5$d$ peaks.
Figures 5(a-e) and 5(f-j) show the experimental and fitted core level spectra of cleaved and S$\&$A samples, respectively, which are obtained using 80-eV photons.
Asymmetric pseudo-Voigt function \cite {Schmid15} and Shirley-type background were used for the fitting.
The magnitude of spin-orbit splitting in Sb 4$d$, Ga 3$d$, and Bi 5$d$ was set to 1.25, 0.43 and 3.06 eV, respectively, according to earlier studies \cite {Eastman80,Gemmeren95}.
Table II summarizes the ratios of peak areas between different elements obtained from the spectra in Figs. 4 and 5.
The peak areas of each element are obtained by integrating the spectra around the target peak after background subtraction in Fig. 4.
The peak-area ratios between each species, such as Ga-S and Ga-B, are obtained from the fitting results shown in Fig. 5.
The fittings were performed with the minimum number of parameters to reproduce the measured spectra, except for the Bi 5$d$ levels on the S$\&$A substrate; the details are provided as follows.

Table II shows the peak-area ratios between different elements.
The obtained ratio is $\sim$1.96 (1.92) for cleaved (S$\&$A) GaSb(110) from Figs. 4(b,c) and 4(d,f).
Figures 5(a) and 5(b) show the core-level spectra of Sb 4$d$ and Ga 3$d$ for the cleaved GaSb(110) substrate and their fittings.
The peaks in the examined area ``B'' (shaded area ``S'') correspond to the contribution from the bulk (surface) Sb and Ga orbitals, which are similar to those shown in Fig. 4.
These assignments are the same as those of the earlier work \cite {Eastman80}.
The surface component S of Ga 3$d$ (Sb 4$d$) shifts toward higher (lower) binding energy side due to the charge transfer between Ga and Sb, which corresponds to their displacement from the original (bulk) position in the topmost atomic layer.
The Sb 4$d$ spectrum in Fig. 5(f) is split into two components, similar to the cleaved substrate in Fig. 5(a), but the peak intensity from the surface contribution S decreases.
Moreover, a new feature S' on Ga 3$d$ in Fig. 5(g) appears at a lower binding energy than the bulk (B) and surface (S) components.
This feature can be attributed to the non-bonding states of Ga atoms, similar to that observed in the core-level spectrum of In 4$d$ for S$\&$A InAs(110) surface \cite {Martinelli97}.
These results suggest that Sb atoms are preferentially removed from the surface by Ar-ion sputtering, and remnant Ga atoms form a cluster on the surface.
Such element-selective sputtering is natural because of the different Ar sputter yields for each element, Ga and Sb in this case \cite{Gnaser95}.

Figures 5(f-e, h-j) show the core-level spectra of Sb 4$d$, Ga 3$d$, and Bi 5$d$ for cleaved and S$\&$A Bi/GaSb.
In the former two, the intensity ratio between S and B are similar in both the substrates, except for the non-bonding feature S' in Ga 3$d$.
The Bi 5$d$ levels of cleaved Bi/GaSb (Fig. 5(d)) could be decomposed into only three components from the current dataset.
The highest (lowest) binding energy component ${\rm Bi_a}$ (${\rm Bi_c}$) is mainly induced by Bi-Sb (Bi-Ga) bonding according to earlier reports \cite {Gemmeren95, Gemmeren98}.
The origin of the middle one (${\rm Bi_b}$) is uncertain; this peak may be induced by the overlap of both Bi atoms, which are bonded with anions and cations.
From the Bi-(2$\times$1) structure (see Fig. 1), Bi atoms should occupy four different sites.
However, we could not decompose four different peaks from the current data, possibly due to the limited energy resolution.
On the other hand, it is difficult to fit the Bi 5$d$ levels of S$\&$A Bi/GaSb with the same number of components as those used in the fitting of cleaved Bi/GaSb.
As the band dispersions of surface states are same for both the substrates (Figs. 2 and 3), it can be assumed that the Bi-(2$\times$1) surface atomic structure is similar for both of them.
Based on this assumption, we added one more peak (${\rm Bi_d}$) for fitting the curve corresponding to S$\&$A Bi/GaSb.
For the other three peaks (${\rm Bi_a}$ -- ${\rm Bi_c}$), the relative intensity and energy differences were kept at the same values as those for the cleaved substrate.
Using this method, we could reproduce the Bi 5$d$ levels of S$\&$A Bi/GaSb, which is shown in Fig. 5(i).
The ${\rm Bi_d}$ peak lies between ${\rm Bi_b}$ and ${\rm Bi_c}$.
After the formation of the Bi-(2$\times$1) surface, the peak-area ratios corresponding to Sb/Ga and Bi/Sb are obtained as $\sim$2.27 and 0.54 for cleaved Bi/GaSb ($\sim$2.17 and 0.89 for S$\&$A Bi/GaSb), respectively.
This implies that S$\&$A Bi/GaSb contains larger amount of Bi than cleaved Bi/GaSb.
The additional ${\rm Bi_d}$ component may be attributed to the insertion of Bi atoms in the Sb vacancies formed by Ar-ion sputtering.

As summarized in Tab. III, all peak widths of the S$\&$A samples are wider than those of the cleaved ones.
For the fitting, we fixed the widths of the surface peaks to be the same as those of the corresponding bulk peaks in the same elements in each same surface treatment, except for ${\rm Bi_d}$ discussed above.
Such peak broadening is the most significant in the Ga peaks, possibly because some Ga atoms no longer have its bonding partner (Sb) after Ar-ion sputtering.
These peak broadenings are consistent with the slightly broad ARPES images and LEED patterns of the S$\&$A surfaces (see Figs. 1 and 2).

Based on the above analysis, there are two possible sources for the hole doping in the surface bands.
The first one is the formation of surface Ga clusters observed as the S' features in Ga 3$d$ levels on the S$\&$A substrate (Fig. 5(j)).
This Ga cluster can attract surface electrons, thereby causing hole doping into the surface Bi Q1D chains.
The second probable origin is the insertion of Bi in the Sb vacancy sites on the GaSb substrate.
The electronegativity of Bi is 2.0, which is equal to that of Sb.
Therefore, from the viewpoint of electronegativity, substituted Bi cannot contribute to the hole doping.
However, as the ionic radius of Bi is larger than that of Sb, slight distortion and/or chemical pressure in the surface atomic structure can be induced.
Such structural deformation can contribute to hole doping, e.g., through Jahn-Teller effect.
It should be noted that the surface band dispersions are not sensitive to such mechanisms and their variations are negligible, similar to that observed for rigid energy shifts.
This suggests that the Bi zig-zag chains formed in the topmost surface layer are an essential part of the Rashba-split surface bands, and the hole-doping mechanism is not related to them, occurring in the sub-surface region.

It should be noted that the similar energy shifts also occur in the core-level spectra, as summarized in Tab. IV.
Both in the GaSb(110) substrates and Bi/GaSb(110)-(2$\times$1) surfaces, all the core-level peaks shift to the lower binding energies by S$\&$A.
However, the core-level energy shifts are 60$\pm$20 meV, which are smaller than those observed in the surface valence bands.
It suggests that the hole doping mainly occurs in the valence bands of surface Bi atoms and the shifts of the other states occur by dragging like the chemical shifts in the core-levels caused by the valence-electron modifications.

\section{Summary}
In this work, we fabricated the surface electronic states with giant Rashba-type spin splitting on Bi/GaSb(110)-(2$\times$1) and succeeded to tune its Fermi level to cause the in-situ transition from semiconductor to surface metal with spin-polarized conduction path.
The surface band of Bi/GaSb(110) exhibits quasi 1D feature with Rashba parameter $\alpha _{\rm R}$ = 2.6 -- 4.1 eV\AA.
The Fermi level of the surface states of Bi/GaSb(110) was tuned by element-selective Ar-ion sputtering on the GaSb substrate.
The surface bands shifted to the lower binding energies rigidly by $\sim$95 meV, which indicates the presence of metallic conduction paths on the surface.
As this metallic and spin-polarized surface is fabricated in-situ by Ar-ion sputtering, it would be easy to design the miniaturized surface circuits by using focused ion-beam method.
Such system would provides a useful platform for future spin-dependent energy conversions such as spin-charge conversion and spin-current-induced terahertz emission.

\section{Acknowledgement}
We acknowledge F. Deschamps for her support during the experiments on the CASSIOP\'EE beamline at synchrotron SOLEIL.
A part of the ARPES measurements was performed under the Nanotechnology Platform Program at IMS of the Ministry of Education, Culture, Sports, Science and Technology (MEXT), subject no. S-18-MS-0048.
SARPES experiment in this work was jointly conducted with ISSP, the University of Tokyo. This work was also supported by JSPS KAKENHI (JP17K18757, JP19J21516, JP19H00651, and JP19H01830).

\newpage

\begin{figure}
\includegraphics[width=80mm]{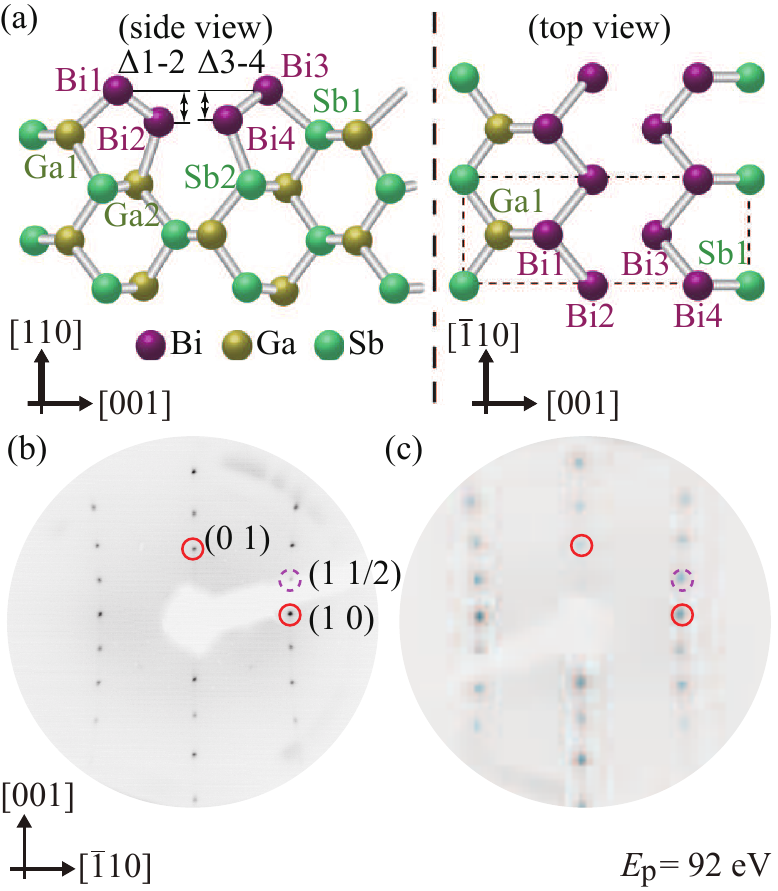}
\caption{\label{figure 1}
(a) Surface atomic structure of Bi/GaSb(110)-(2$\times$1) reported in Ref. \cite {Gemmeren98}.
The dashed rectangle (in the figure on the right) indicates the (2$\times$1) unit cell.
(b, c) LEED patterns of the Bi/GaSb(110)-(2$\times$1) surface, where the substrates were cleaned by (b) cleaving and (c) repeated cycles of Ar-ion sputtering and annealing. Both LEED patterns were obtained at 60 K.}
\end{figure}

\begin{figure*}[p]
\includegraphics[width=150mm]{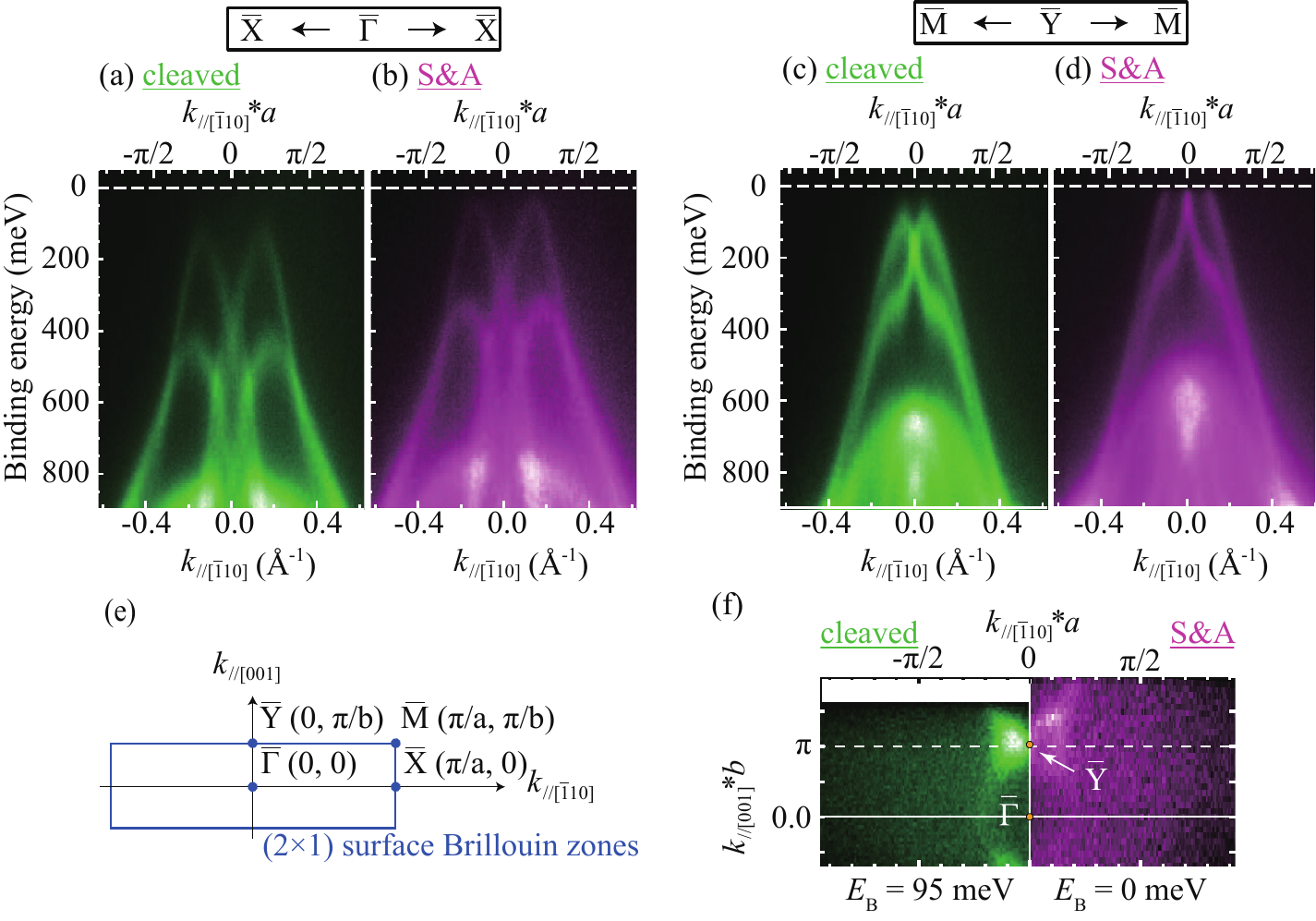}
\caption{\label{figure 2}
(a-d)ARPES intensity plots obtained from cleaved (a, c) and S$\&$A (b, d) Bi/GaSb at 20 K ($h\nu$ = 30 eV) along $\bar{\Gamma}$--$\bar{\rm X}$ ($k_{//[001]}*${\sl b} = 0) (a, c) and $\bar{\rm Y}$--$\bar{\rm M}$ ($k_{//[001]}*${\sl b} = $\pi$) (c, d).
Dashed lines indicate the Fermi level.
(e) (2$\times$1) SBZ.
(f) Constant-energy contours of the cleaved (left) and S$\&$A (right) Bi/GaSb surfaces taken with the energy windows of $\pm$ 20 meV. Binding energies are shown in the figure. The dashed line represents a boundary of the SBZ
}
\end{figure*}

\begin{figure}[p]
\includegraphics[width=80mm]{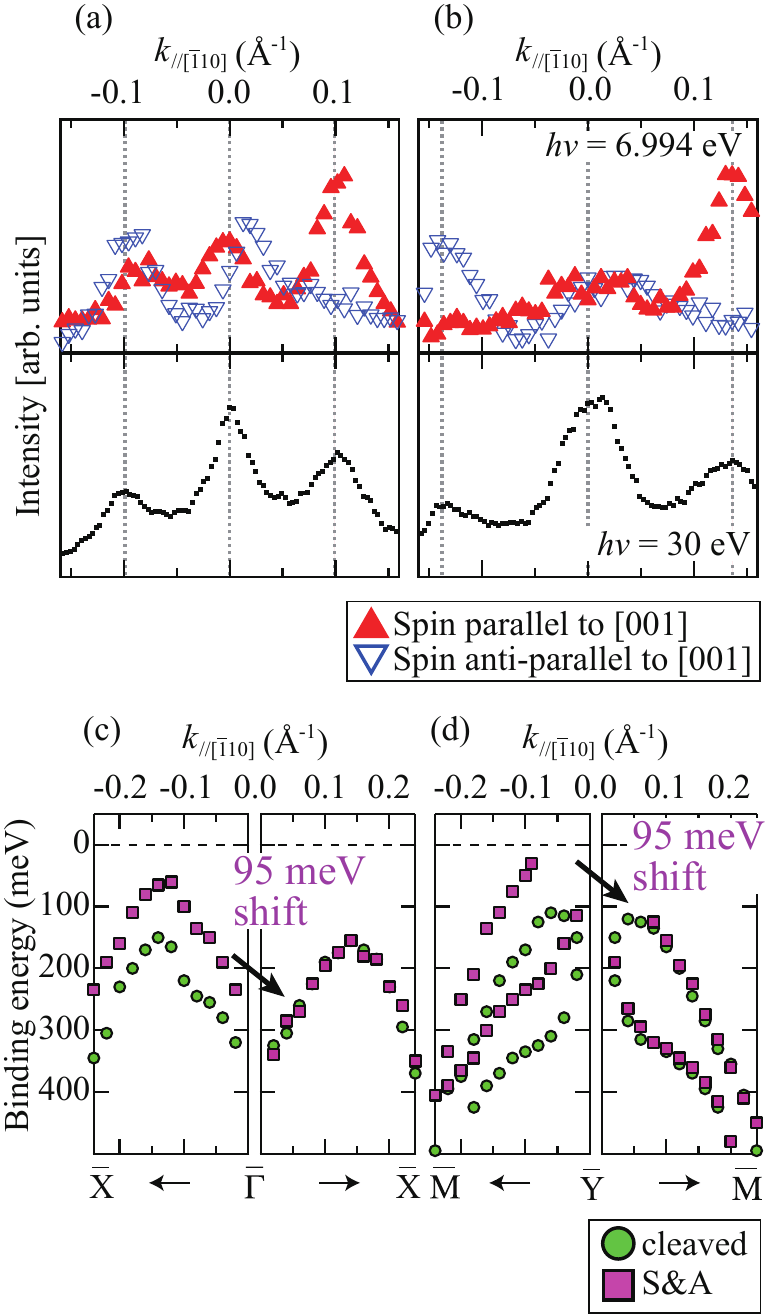}
\caption{\label{figure 3}
(a, b) Spin-resolved (top) and spin-integrated (bottom) momentum distribution curves (MDCs) along $\bar{\rm Y}$--$\bar{\rm M}$ obtained with 6.994-eV and 30-eV photons of S$\&$A Bi/GaSb at 14 K, respectively.
The binding energies for (a) and (b) are 20 meV and 100 meV, respectively.
(c, d) Peak positions of the energy distribution curves along $\bar{\Gamma}$--$\bar{\rm X}$ (c) and $\bar{\rm Y}$--$\bar{\rm M}$ (d) in the left-side figures.
The data corresponding to S$\&$A Bi/GaSb are shifted by 95 meV in the right-side figures.
Circles (squares) represent cleaved (S$\&$A) Bi/GaSb.
}
\end{figure}

\begin{figure}[bp]
\includegraphics[width=80mm]{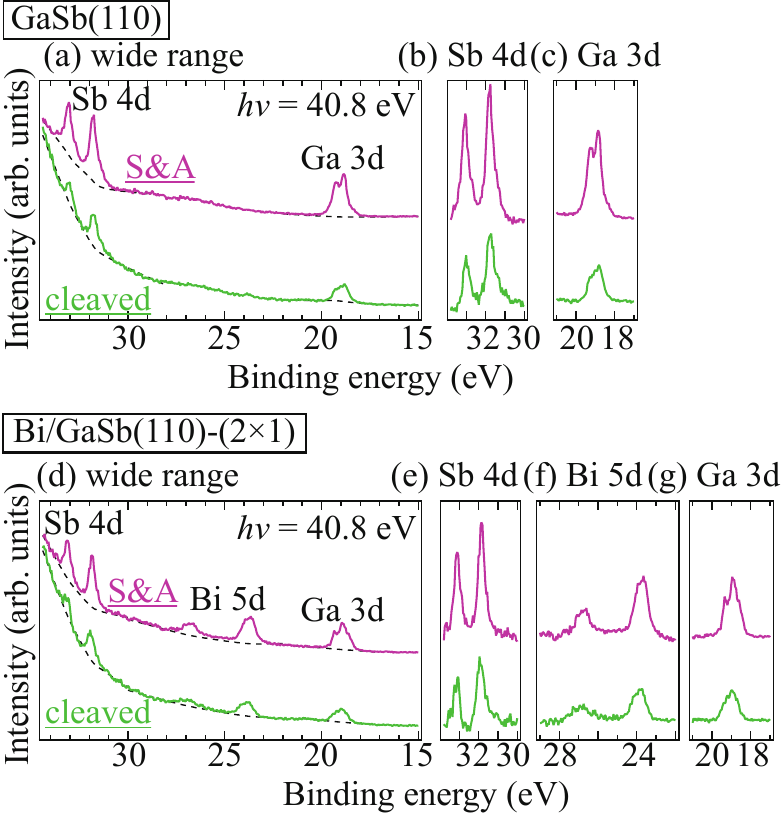}
\caption{\label{figure 4}
Core-level spectra of (a-c) GaSb(110) substrate and (d-g) Bi/GaSb(110)-(2$\times$1) surface at 80 K ($h\nu$ = 40.8 eV).
(a,d) Wide-range spectra with background.
The pink (green) spectra correspond to S$\&$A (cleaved) GaSb(110) substrate.
Dashed lines indicate Shirley-type background.
(b-c,e-g) Core-level spectra obtained after background subtraction from (a,d).
}
\end{figure}

\begin{figure*}[htbp]
\includegraphics[width=150mm]{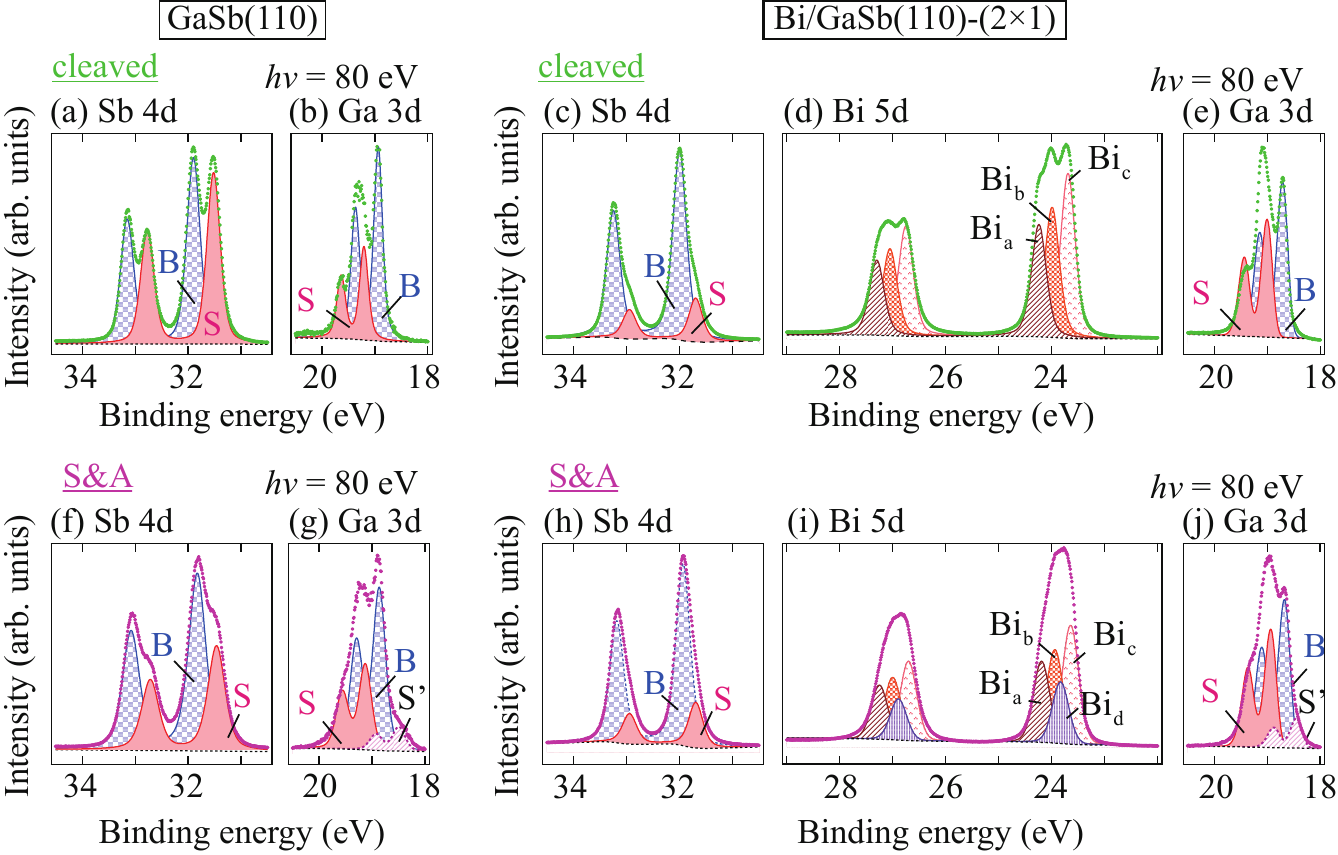}
\caption{\label{figure 5}
Detailed core-level spectra of (a,b,f,g) GaSb(110) substrates and (c-e,h-j) Bi/GaSb(110)-(2$\times$1) surfaces, obtained using 80-eV photons at 14 K.
The top (a-e) and bottom (f-j) rows correspond to the cleaved and S$\&$A samples, respectively.
The green (top row) and pink (bottom row) dotted lines represent the raw spectra fitted using Shirley-type background and pseudo-Voigt function.
}
\end{figure*}

\newpage

\begin{table}
\caption{
Bond length and relative atomic displacements of the Bi-atomic chains along the surface normal (the notations are provided in Fig. 1(a)) \cite {Lottermoser98, Gemmeren98, Gay00, Betti99}.
}
\label{table2}
\begin{ruledtabular}
\begin{tabular}{lcc}
 & \multicolumn{2}{c}{Distance (\AA)} \\
 & Bi/GaSb &Bi/InAs \\
\hline
Bi1-Bi2 & 3.06 &2.98 \\
Bi3-Bi4 & 3.04 & 2.93 \\
Bi1-Ga1 (In1) & 2.79 & 2.62 \\
Bi2-Ga2 (In2) & 2.87 & 2.98 \\
Bi3-Sb1 (As1) & 2.87 & 3.05 \\
Bi4-Sb2 (As2) & 2.96 & 3.10 \\
$\Delta_{1-2}$ & 1.34 & 1.29 \\
$\Delta_{3-4}$ & 1.21 & 1.06
\end{tabular}
\end{ruledtabular}
\end{table}

\begin{table}
\caption{
The ratios of the peak areas between different elements in (a) the GaSb(110) clean substrate and (b) the Bi/GaSb(110)-(2$\times$1) surface obtained from Figs. 4 and 5. The evaluation method is described in the text.
Fractions such as ``Ga/Sb'' and ``Bi$_b$/Bi$_a$'' represent the ratio between the integrated peak areas (Sb 4$d$, Ga 3$d$, and so on).
}
\label{table1}
\begin{flushleft} (a) GaSb(110)-(1$\times$1) \end{flushleft}
\begin{ruledtabular}
\begin{tabular}{lcccc}
& Sb/Ga & Ga S/B & Ga S'/B & Sb S/B \\
\hline
Cleaved & 1.96 & 0.48 & - & 0.92 \\
S$\&$A & 1.92 & 0.53 & 0.15 & 0.60 \\
\end{tabular}
\end{ruledtabular}
\vspace{0.2in}\\
\begin{flushleft} (b) Bi/GaSb(110)-(2$\times$1) \end{flushleft}
\begin{ruledtabular}
\begin{tabular}{lcccccccc}

& Sb/Ga & Bi/Sb & Ga S/B & Ga S'/B & Sb S/B & ${\rm Bi_b}/{\rm Bi_a}$ & ${\rm Bi_c}/{\rm Bi_a}$ & ${\rm Bi_d}/{\rm Bi_a}$ \\
\hline
Cleaved & 2.27 & 0.54 & 0.76 & - & 0.23 & 1.16 & 1.47 & -\\
S$\&$A & 2.17 & 0.89 & 0.80 & 0.20 & 0.25 & 1.16 & 1.47 & 0.76 \\
\end{tabular}
\end{ruledtabular}
\end{table}

\begin{table}
\caption{
The peak widths (full width at half maximum) of the fitting components. All widths are in units of eV.
}
\label{table4}
\begin{ruledtabular}
\begin{tabular}{lccccc}
& \multicolumn{2}{c}{GaSb(110)-(1$\times$1)} & \multicolumn{3}{c}{Bi/GaSb(110)-(2$\times$1)} \\
& Ga & Sb & Ga & Sb & Bi\\
\hline
Cleaved & 0.20 & 0.30 & 0.20 & 0.31 & 0.33 \\
S$\&$A & 0.28 & 0.39 & 0.28 & 0.32 & 0.37 \\
\end{tabular}
\end{ruledtabular}
\end{table}

\begin{table}
\caption{
The relative peak positions of the fitted core-level spectra from (a) the GaSb(110) clean substrates and (b) the Bi/GaSb(110)-(2$\times$1) surfaces. Binding energies of B (for Bi, Bi$_a$) from the cleaved surfaces are shown in the parenthesis. All peak positions are in units of eV.
}
\label{table3}
\begin{flushleft}　(a) GaSb(110)-(1$\times$1)　\end{flushleft}
\begin{ruledtabular}
\begin{tabular}{lcccccc}
&\multicolumn{3}{c}{Ga 3$d$} & \multicolumn{2}{c}{Sb 4$d$} \\
& B & S & S' & B & S  \\
\hline
cleaved & 0 (18.94) & $+$0.27 & - & 0 (31.90) & $-$0.38 \\
S$\&$A  & $-$0.08 & $+$0.19 & $-$0.44 & $-$0.07 & $-$0.43  \\
\end{tabular}
\end{ruledtabular}
\vspace{0.2in}\\
\begin{flushleft}　(b) Bi/GaSb(110)-(2$\times$1)　\end{flushleft}
\begin{ruledtabular}
\begin{tabular}{lccccccccc}
&\multicolumn{3}{c}{Ga 3$d$} & \multicolumn{2}{c}{Sb 4$d$} & \multicolumn{4}{c}{Bi 5$d$}  \\
 & B & S & S' & B & S & Bi$_a$ & Bi$_b$ & Bi$_c$ & Bi$_d$\\
\hline
cleaved & 0 (18.72) & $+$0.29 & - & 0 (32.00) & $-$0.31 & 0 (24.24) & $-$0.25 & $-$0.54 & - \\
S$\&$A & $-$0.04 & $+$0.22 & $-$0.27 & $-$0.07 & $-$0.31 & $-$0.05 & $-$0.30 & $-$0.60 & $-$0.45 \\
\end{tabular}
\end{ruledtabular}
\end{table}

\end{document}